\documentclass[12pt]{iopart}

\usepackage{cite}
\usepackage{amssymb}
\newcommand{\sech}{\mathop{\mathrm{sech}}\nolimits}
\usepackage[mathscr]{euscript}
\usepackage{bm}
\usepackage{braket}
\usepackage{color}

\usepackage{graphicx}
\usepackage[hidelinks]{hyperref}

\begin{document}

\title{The Globally Trapped Future: A Fate for Black Holes and Wormholes}

\author{Yi-Bo Liang$^1$ and Hong-Rong Li$^1$}

\address{$^1$School of Physics, Xi’an Jiaotong University, Xi’an 710049, China}

\ead{liangyibo@stu.xjtu.edu.cn, hrli@xjtu.edu.cn}

\vspace{10pt}

\begin{abstract}
We demonstrate, for the first time, that arbitrary spherically symmetric metrics can be derived within a framework based on the coupling of two scalar fields and an electromagnetic field. We then specialize to a class of non-stationary spacetimes characterized by analytically tractable global causal structure and trapping horizons, which is particularly suited for investigating the evolution of black holes and wormholes. Within this framework, we find that the fate of spacetime can be categorized into three distinct classes: those without a globally trapped future; those without a globally trapped future but containing bounded, Cauchy-foliated trapped regions; and those with a future that becomes completely trapped. The evolution of a geometrically Schwarzschild-like black hole and a horizonless wormhole demonstrates these possible fates, thus revealing the globally trapped outcome as a novel theoretical possibility. Finally, we propose that a regular black hole may contain an unattainable minimal throat---a minimal-radius throat that is causally unreachable---and refer to it henceforth as the Endless Throat.

\end{abstract}

	
	\section{Introduction}
	Black holes are not only a cornerstone prediction of Einstein's gravity but also an established astrophysical reality.
	The Schwarzschild black hole \cite{schwarzschild1916gravitationsfeld,wald2024general} and its distinct bifurcate horizon provide the essential framework for black holes in general relativity; nevertheless, it possesses a spacetime singularity \cite{penrose1965gravitational,hawking1967occurrence,hawking1970singularities} that indicates the theory's breakdown, thereby preventing classical predictions, information retrieval, or a smooth extension of the spacetime.
	Contemporary research extends this paradigm to eliminate singularities, leading to the concept of regular black holes \cite{bardeen1968non,ayon1998regular,ayon2000bardeen,bronnikov2000comment,bronnikov2001regular,burinskii2002new,fan2016construction,rodrigues2022bardeen,dariescu2022kiselev,bronnikov2023regular,moreno2003stability,novello2000geometrical,novello2000singularities,ma2014corrected,bokulic2021black,simpson2019black,lobo2021novel,bronnikov2022black,bronnikov2022field,canate2022black}—horizoned objects devoid of a central singularity.
	This raises a fundamental question: how do such non-singular objects evolve dynamically?
	
	Influential models and analyses \cite{hayward1999dynamic,shinkai2002fate,hayward2009wormhole,bronnikov2007regular,Hayward:2005gi,simpson2019vaidya,lobo2020dynamic,yang2021trapping} suggest that a black hole may theoretically transition into a traversable wormhole, and vice versa.
	This framework enables a re-examination of the wormhole-black hole relationship by analyzing transformations in the causal structure of their trapping horizons.
	Indeed, the Schwarzschild black hole can be reinterpreted as a wormhole with unidirectional traversability, where the central singularity blocks all passage.
	Scientific inquiry extends beyond the mere interconversion of these objects to encompass their ultimate fate.
	A pivotal question is whether, within a finite time, they will eventually engulf the entire future spacetime.
	However, obtaining exact solutions for non-stationary (dynamical) evolution in gravitational systems is inherently difficult, hence progress in this line of research has been slow.
	
	The primary aim of this paper is to construct and employ a novel framework for generating spherically symmetric metrics, built from two scalar fields coupled to an electromagnetic field. When applied to analytically manageable non-stationary spacetimes, this framework leads to a three-fold classification of their ultimate evolution. Beyond mere taxonomy, this classification—demonstrated through dynamical studies of benchmark cases—establishes the globally trapped scenario as a compelling new possibility.
	
	\section{Theoretical Framework}
	An arbitrary spherically symmetric metric can be expressed in the general local form:
	\begin{equation}
		ds^2=-A(T,X)\mathrm{d}T^2+A(T,X)\mathrm{d}X^2+r^2(T,X)\mathrm{d}\Omega^2.
	\end{equation}
	In fact, it is an exact solution deriving from an action
	\begin{eqnarray}
		S = \int_M \epsilon \Big[ & R- p_1\, \nabla^a \phi_1 \nabla_a \phi_1- p_2\, \nabla^a \phi_2 \nabla_a \phi_2 \nonumber\\
		& - p_3\, \nabla^a \phi_1 \nabla_a \phi_2- p_4 \, \mathscr{L}(F) \Big],\label{2}
	\end{eqnarray}
	that comprises two scalar fields and an electromagnetic field ($c=G=4\pi\varepsilon_0=1$).
	Here, the functions $p_1,p_2,p_3,p_4$ all depend on the scalar fields $\phi_1$ and $\phi_2$.
	Parameters $p_1,p_2$ and $p_3$ encode the parametrization freedom of the scalar fields and their kinetic couplings.
	A non-positive-definite kinetic matrix signals the presence of phantom energy, which is absent if the matrix is positive-definite.
	This treatment for extracting the parametrization freedom is inspired by \cite{bronnikov2022black}.
	In the electromagnetic sector, we focus on a magnetic field characterized by $\mathscr{L}(F)=F=F_{ab}F^{ab}=2g^2/r^4$ ($g$ being the magnetic charge).
	The coupling between $F$ and the scalars $\phi_1, \phi_2$ is mediated through $p_4$.
	The solution (exploiting reparametrization freedom by setting $\phi_1\equiv X,\phi_2\equiv T$) is
	\begin{eqnarray}
		p_1=G_{TT}-\frac{A}{r^2}G_{\theta\theta},\ p_2=G_{XX}+\frac{A}{r^2}G_{\theta\theta},\\
		p_3=2G_{TX},\ p_4=\frac{1}{A F}\left(G_{TT}-G_{XX}\right),
	\end{eqnarray}
	where $G_{ab}$ is the Einstein tensor.
	With $\phi_1= X,\phi_2= T$, the functions $p_1,p_2,p_3,p_4$ and $V$ can be expressed simply in terms of $\phi_1,\phi_2$.
	Then, the equations of motion for $\phi_1$ and $\phi_2$ can be checked to verify that they are satisfied.
	Within our theoretical setup, the standard source descriptions of the Schwarzschild black hole and the Ellis wormhole are naturally recovered.
	In contrast, (anti-)de Sitter spacetime — which is sourced by a cosmological constant — has a different source structure, as the potential term $V(\phi_1,\phi_2)$ is absent in Eq.~\eref{2}.
	The corresponding metric components are $A=1-\tanh^2(\sqrt{\Lambda/3}X)$ and $r=\tanh(\sqrt{\Lambda/3}X)/\sqrt{\Lambda/3}$.
	Its source terms are given by
	\begin{equation}
		p_1=-p_2=2\Lambda \sech^2(\sqrt{\Lambda/3}X),\ 
		p_3=0,\ p_4=\frac{9\tanh^4(\sqrt{\Lambda/3}X)}{\Lambda g^2},
	\end{equation}
	where $p_1$, $p_2$, and $p_4$ tend to zero as $\Lambda \to 0$.
	
	In our analysis of black holes and wormholes, we adopt three key restrictions derived from the intrinsic properties of the maximally extended Schwarzschild black hole (MESBH).
	First, we assume a diagonal energy-momentum tensor for the spacetime, which greatly simplifies the analysis of energy conditions.
	Then, we have
	\begin{equation}
		\partial_{T} A \partial_{X} r +\partial_{X} A \partial_{T} r-2 A \partial_{T,X}r=0.\label{7}
	\end{equation}
	A sufficient condition for the solvability of the equation for a general $r$, namely the separability of the derivative ratio $\partial_{T} r/\partial_{X} r=P(T)/Q(X)$, leads to the general solution $r=R(f(X)+h(T))$.
	Therefore, Eq.~\eref{7} can be solved to yield
	\begin{equation}
		A=F(f-h)R^\prime(f+h),\ r=R(f+h),\label{8}
	\end{equation}
	where $R^\prime\equiv dR/d(f+h)$.
	Clearly, for the MESBH, we have $J(r)\equiv(r/(2M)-1)e^{r/(2M)}$, $f=X^2$, $h=-T^2$, and $dJ/dr> 0$. Since $J(r)=f+h$, it follows that $r=J^{-1}(f+h)\equiv R(f+h)$.
	Hence, we secondly require the smooth functions $f = f(X^2)$ and $h = h(T^2)$ to preserve both the time-reversal symmetry and the isometry $X \to -X$, with the additional condition $\partial_T h < 0$ for $T>0$.
	Third, we assume globally defined coordinates $\{T,X,\theta,\phi\}$ ($A>0, r>0$)—where $T\to \infty$ and $X\to \infty$ correspond to infinities—along with a globally hyperbolic spacetime of topology $\mathbb{R}^2\times \mathbb{S}^2$, all properties inherent to the MESBH.
	
	The two radial null geodesic tangent vectors are defined as
	$k_{\pm}^a\equiv (1/\sqrt{2A})((\partial/\partial T)^a\pm(\partial/\partial X)^a)$.
	Their expansions, $\theta_{k_{\pm}}=h^{ab}\nabla_a k_{\pm b}$, are computed using the induced metric $h_{ab}\equiv g_{ab}+k_{+a} k_{-b}+k_{-a} k_{+b}$, yielding the explicit expressions: $\theta_{k_{\pm}}=\sqrt{2/A}(R^\prime/R)(\partial_T h\pm\partial_X f)$.
	If the metric is regular, since $F, R^\prime$ are functions of independent variables, they must be nonzero and thus possess definite signs.
	Without loss of generality, we set $R^\prime>0$, because the case $R^\prime<0$ can be transformed into the former by defining $R(f+h)=R(-(-f-h))\equiv\tilde{R}(\tilde{f}+\tilde{h})$ with $\tilde{f}\equiv -f,\tilde{h}\equiv -h$, ensuring $\tilde{R}^\prime>0$.
	Consequently, we have $\mathrm{sign}[\theta_{k_\pm}]=\mathrm{sign}[\partial_T h\pm\partial_X f]$.
	The trapped regions are precisely those where $\partial_T h\pm\partial_X f< 0$.
	With $\partial_T h < 0$ for $T>0$ (leads to a bounce in the constant-$X$ slices), we can establish several general properties:
	
	(i) The spacetime will not be entirely trapped in the future if $\sup|\partial_X f| =+\infty$ or if $\sup|\partial_X f| =\delta$ provided that $\displaystyle\lim_{T \to +\infty} \partial_T h\geq-\delta$.
	
	(ii) Any spacetime region defined by $0<T_0 < T < T_1$ and satisfying $\sup|\partial_X f| \leq \delta$ and $\partial_T h \leq -\delta$ is trapped.
	It follows that one or more such trapped regions can exist.
	
	(iii) The remaining scenario is characterized by the asymptotic conditions $\partial_T h < -\delta$ as $T \to +\infty$ and $\sup|\partial_X f| \leq \delta$.
	Integration of the condition on $h$ implies $h(T^2)<h(T^2_0)-\delta(T-T_0)$, that is, $\displaystyle\lim_{T \to +\infty}h=-\infty$.
	Thus, for constant $X$, the variable $x \equiv f + h$ monotonically decreases to $T\to-\infty$.
	Given $R > 0$ and $R^\prime > 0$, $R(x)$ must approach its infimum $L \geq 0$ as $x \to -\infty$.
	Similarly, it can be verified that $R^\prime \to 0^+$ as $x \to -\infty$.
	Therefore, if the spacetime is entirely future-trapped, the areal radius $R$ approaches its minimum value $L$ as $T \to \infty$.
	The spacetime precludes the definition of conformal spacelike boundaries at temporal infinity ($T\to\infty$), and thus, due to the condition $g_{TT}=-g_{XX}$, also precludes conformal timelike boundaries at spatial infinity ($X\to\infty$).
	Introducing the light-cone coordinate $u\equiv T-X$, we observe that $x\to x_0$ or $x\to-\infty$ (since $\partial_T x=\partial_X f+\partial_T h<0$) as $T\to+\infty$ along lines of constant $u$, therefore, $R\to R(x_0)\geq L$.
	Consequently, conformal null boundaries cannot be defined for this spacetime.
	Therefore, this spacetime cannot be asymptotically flat.

	We thus identify three possible fates for the spacetime.
	Clearly, the results given above can be generalized to spacetimes satisfying both the last two restrictions and the condition $r=R(f+h)$.
	
	\section{Concrete Models}
	An initially geometrically Schwarzschild-like black hole (GSLBH) is defined by the condition that its trapping horizons are bifurcated and null at the point $T=X=0$.
	We first construct the simplest regular spacetime that exhibits this behavior and combines the first two scenarios.
	This leads to the form $R=f+h$ ($R^{\prime}=1,F\equiv1$), and the metric is given by
	\begin{equation}
		ds^2=-\mathrm{d}T^2+\mathrm{d}X^2+\left(\sqrt{X^2+a^2}+\frac{m}{1+kT^2}\right)^2\mathrm{d}\Omega^2,
	\end{equation}
	where $m, k\geq0,a>0$.
	The spacetime has finite curvature invariants $R,R_{ab}R^{ab},R_{abcd}R^{abcd}$ and reduces to Ellis–Bronnikov wormhole ($p_2,p_3, p_4\to0$) when $m\to0$ or $T \to \infty$.
	Since the spacetime is asymptotically flat at null infinities \cite{ashtekar2014geometry} and the vector field $(\partial/\partial T)^a$ is asymptotically Killing, it follows that the Bondi energy vanishes.
	The Misner–Sharp mass, $2M = R(1 - (\partial_T h - \partial_X f)(\partial_T h + \partial_X f))$, is used to define the black hole/wormhole mass at the trapping horizon $\mathcal{H}$, leading to the result $2M_{\mathcal{H}} = R_{\mathcal{H}}$.
	
	\begin{figure}[t]      
		\centering             
		\includegraphics[width=0.55\textwidth]{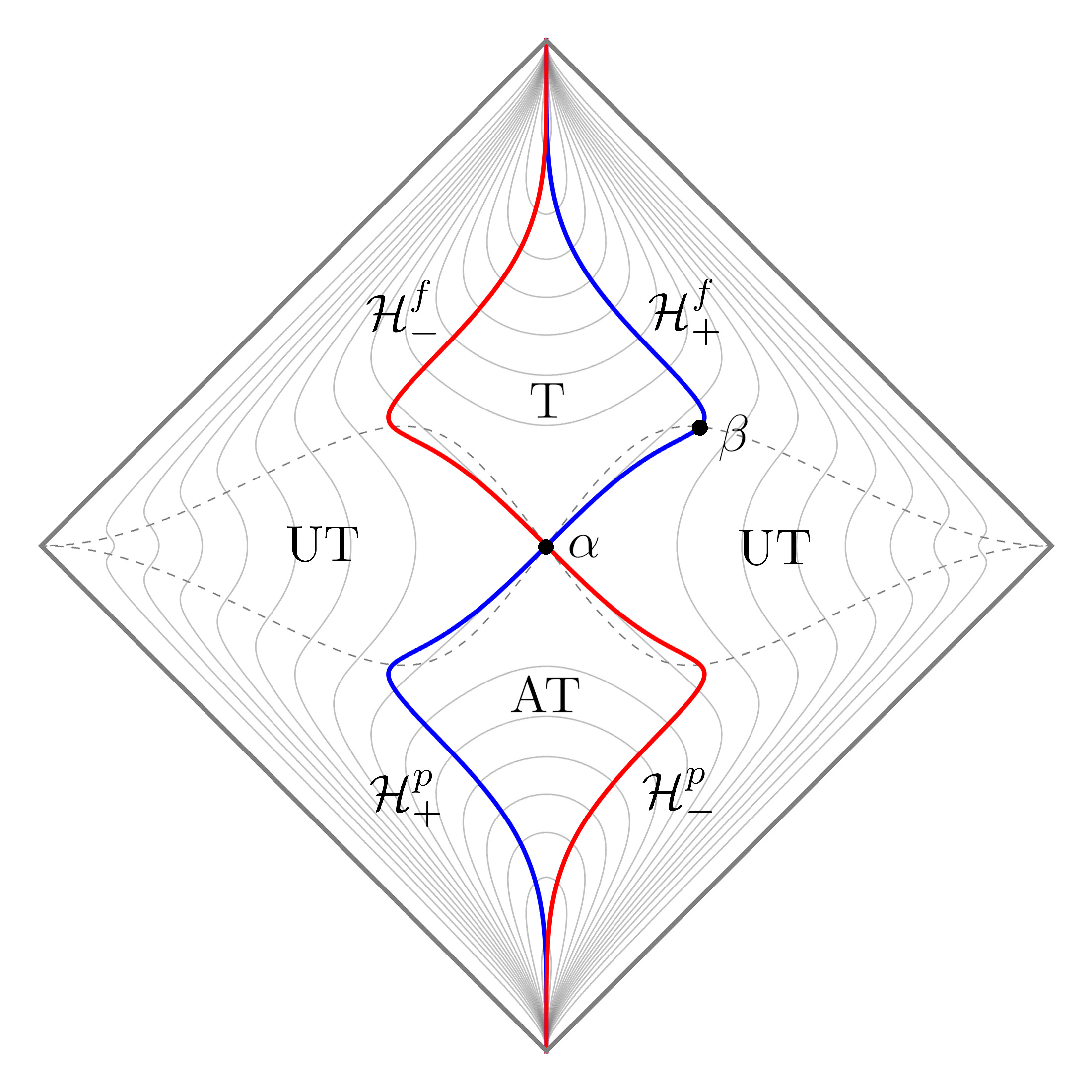}
		\caption{The Penrose diagram for parameters $a=0.4, m=1.4, k=1/(2am)$. Regions: $\mathrm{T}$ (trapped), $\mathrm{AT}$ (anti-trapped), $\mathrm{UT}$ (untrapped). Blue and red lines indicate marginally trapping horizons of $k^a_+$ and $k^a_-$, thick and thin gray lines denote null infinities and constant $R$ surfaces (larger $R$ closer to null infinity). The null energy condition (NEC) and the strong energy condition (SEC) are satisfied in the region between the two dashed lines.}
		\label{f1}    
	\end{figure}
	
	\begin{figure}[htbp]      
		\centering             
		\includegraphics[width=0.55\textwidth]{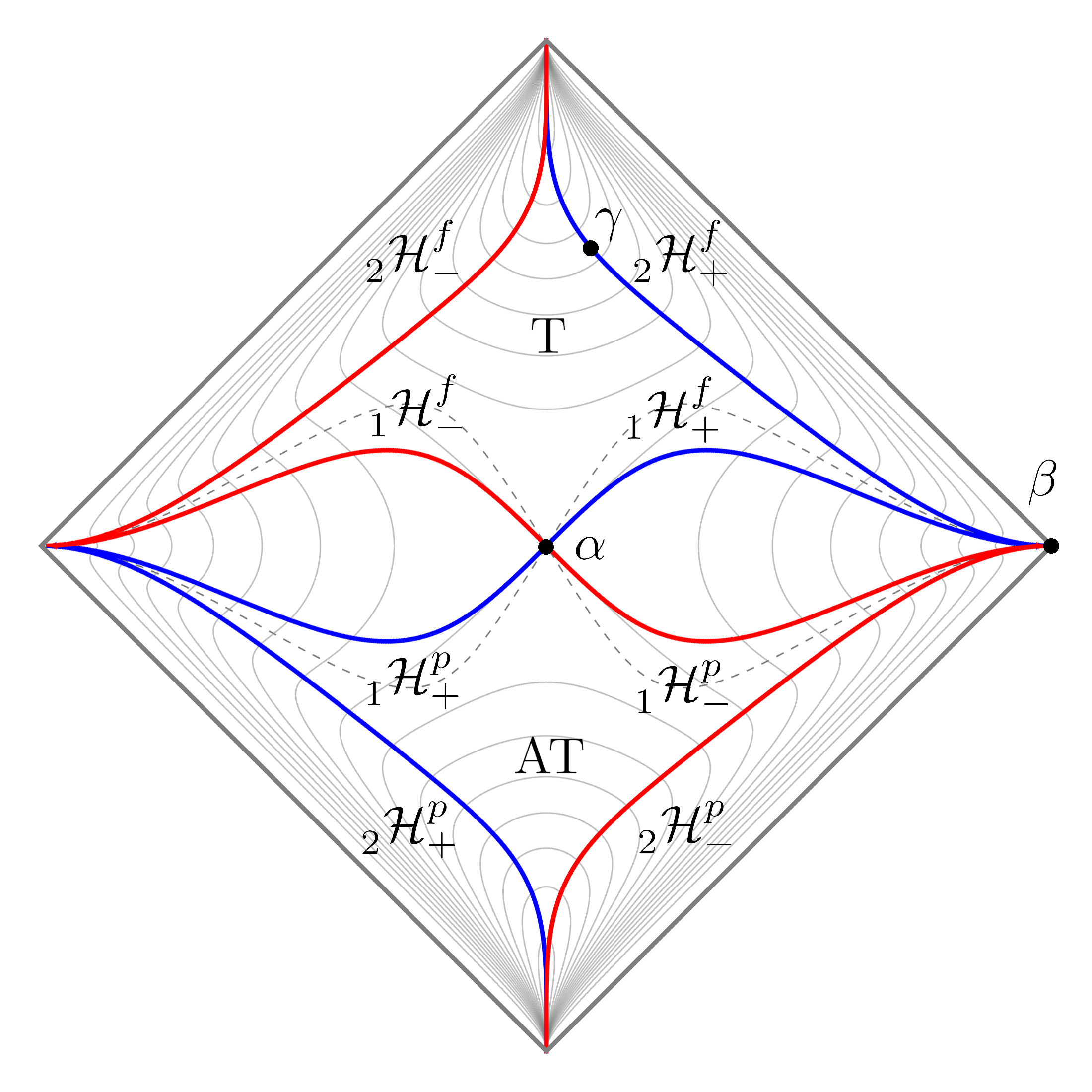}
		\caption{Penrose diagram for $a=0.4, m=2, k=1/(2am)$. $\rho\geq0$ inside the regions enclosed by the four dotted lines. The label $\mathrm{UT}$ is not shown. The NEC and the SEC are satisfied in the region between the two dashed lines.}
		\label{f2}    
	\end{figure}
	
	The throat measures $a+m$ at $T=X=0$, with $a$ representing its global minimum.
	The causal nature of the marginally trapping horizon is determined by its dual normal covector $n_a\equiv (d\theta_{k_+})_a$, with analysis confined to $k^a_+$ by symmetry.
	For $T>0$, the horizon is null at no more than three points, while it becomes timelike as $T \to +\infty$.
	From $\partial_T h(T^2)+\partial_X f(X^2)= 0$, as $T,X\to0^+$, the approximation $X/a+O(X^3)=2mkT+O(T^{3})$ implies that the horizon's character (null, spacelike, or timelike) in this limit is dictated by the value of $mk$.
	
	The special case $2mk = 1/a$ yields a GSLBH as $T,X \to 0^+$; given $\sup \partial_X f = 1$, a Cauchy (surface)-foliated region is trapped if $\inf \partial_T h < -1$, which is equivalent to $m^2k > (4/3)^3$.
	The Penrose–Carter diagram illustrating the evolution of the GSLBH is shown in Figs.~\ref{f1} and~\ref{f2}.
	In Fig.~\ref{f1}, the horizons $\mathcal{H}_+^f$ initially expands and subsequently contracts.
	The surface $\mathcal{H}^f_+$ is a future outer trapping horizon (FOTH) that is null at $\alpha,\beta$, spacelike in $(\alpha,\beta)$, and timelike elsewhere; thus, its one-way traversable spacelike segment (with positive flux into it) defines a black hole horizon.
	The remaining timelike portion of $\mathcal{H}^f_+$ (with negative flux into it) is bidirectional traversable, corresponding to a wormhole horizon.
	The transformation $T\to-T,X\to-X$ clarifies the properties of $\mathcal{H}^p_\pm$ and $\mathcal{H}^f_-$.
	In summary, the spacetime---initially containing a GSLBH---first undergoes accretion, then experiences a classical analogue of evaporation, and finally makes a definitive transition into a traversable wormhole.
	The collections of the throats and bounces are the regions $\mathrm{T} \cup \mathrm{AT}$ and $\mathrm{UT}$, respectively.

	As shown in Fig.~\ref{f2}, a black hole first traps a Cauchy-foliated region and subsequently evaporates into a wormhole.
	This trapped region is bounded by the horizons ${}^{}_{1}\mathcal{H}^f_+,{}^{}_{2}\mathcal{H}^f_+,{}^{}_{1}\mathcal{H}^f_-$ and ${}^{}_{2}\mathcal{H}^f_-$.
	The spacelike segment $ (\beta, \gamma) $ of ${}^{}_{2}\mathcal{H}^f+$—which is null at $ \gamma $—is a future inner trapping horizon (FITP) (Cauchy horizon, with negative flux from it).
	
	To obtain an evolution of an initialy horizonless wormhole, just replace of $kT^2$ in $h$ with $lT^4$.
	Here, horizonless indicates that, at $T=X=0$ and for fixed $\theta,\varphi$, the tangent vectors of the two bifurcated trapping horizons overlap and is timelike.
	A Cauchy-foliated region is trapped if $m l^{1/4}> 2^43^{-3/4}5^{-5/4}$.
	For brevity, the corresponding Penrose-Carter diagrams have been omitted.
	
	\begin{figure}[t]      
		\centering             
		\includegraphics[width=0.55\textwidth]{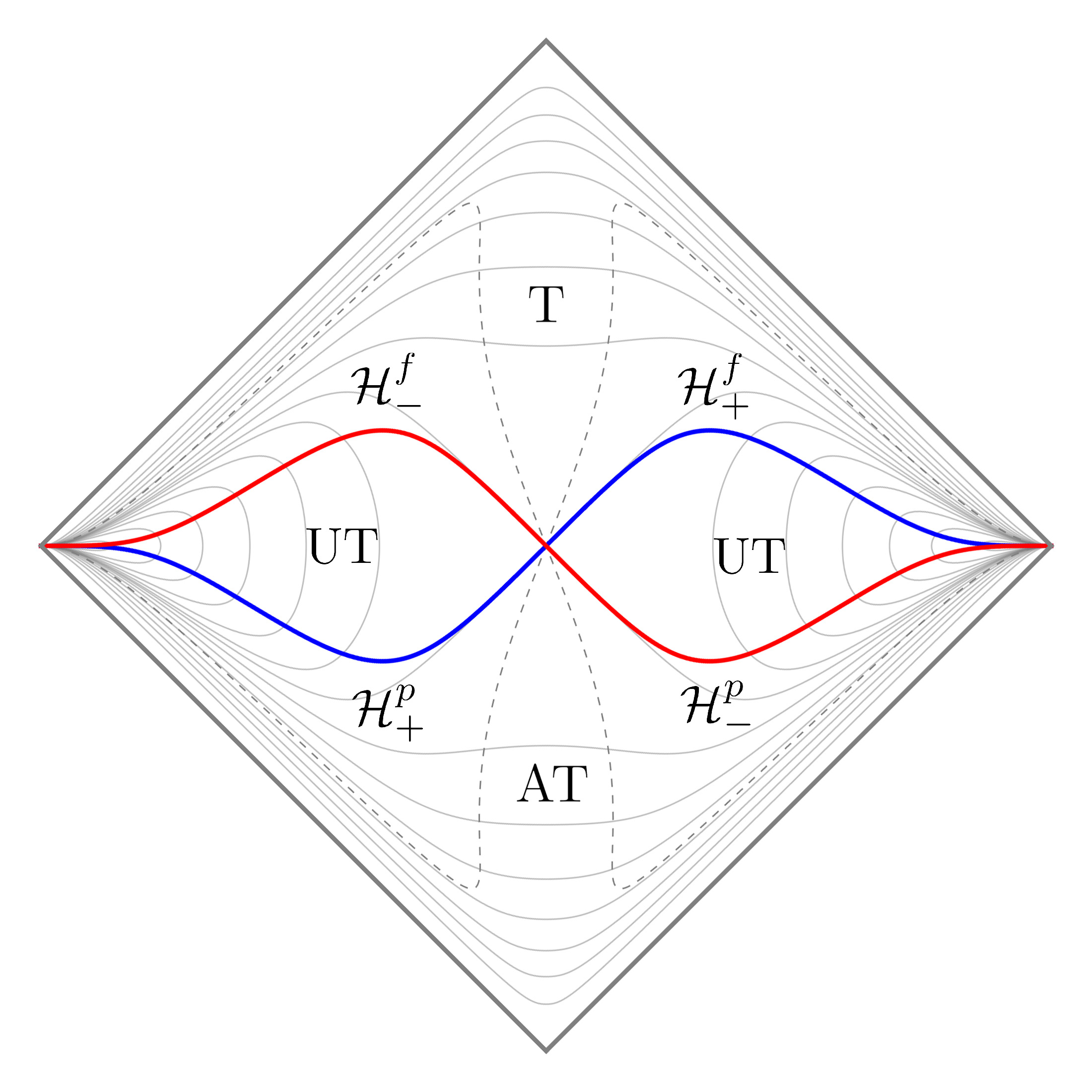}
		\caption{Penrose diagram for $a=2, L=1, k=1/(2a(a-L)^2), mk=4, n=0$. Note that the thick gray lines do not represent conformal boundaries but merely coordinate boundaries, while the thin gray lines indicate surfaces of constant $R$; closer to the boundaries, these surfaces correspond to smaller $R$. The weak energy condition (WEC) is satisfied in the region between the two dashed lines.}
		\label{f4}    
	\end{figure}
	
	Finally, we present a spacetime that possesses a globally trapped future.
	Given the conditions that $h \to -\infty$ as $T \to \infty$ with finite $\partial_X f$, we can explicitly construct the metric as $f=-1/(\sqrt{X^2+a^2}-L),h=-kT^2-mk,R=L-1/(f+h), F=(f-h)^2$ then yields
	\begin{equation}
		ds^2=\left(\frac{f-h}{f+h}\right)^2(-\mathrm{d}T^2+\mathrm{d}X^2)+\left(L-\frac{1}{f+h}\right)^2\mathrm{d}\Omega^2,\label{gtf}
	\end{equation}
	with $a>L>0$ and $m,k\geq0$.
	This spacetime reduces to the Ellis–Bronnikov wormhole for $k=0$, and reduces to a tube-like spacetime with $A\to 1,R\to L$, in the limit $T \to \infty$.
	
	Let us define the throat size at $T=X=0$ as $R_c = L + 1/((a-L)^{-1} + mk)$, its infimum as $R_i = L$, and its supremum as $R_s = L + 1/mk$ (attained at $T=0$ and as $X\to\infty$).
	Furthermore, the regularity condition $A \neq 0$ requires $mk> 1/(a-L)$ for non-zero $k$.
	This leads to a constraint $R_c > (R_i + R_s)/2$.
	One finds directly that $n_a$ is timelike in the limit $X \to \infty$, implying that the horizon $\mathcal{H}^f_+$ is spacelike there.
	Moreover, since the black hole mass satisfies $2M_{\mathcal{H}} = R_{\mathcal{H}}$, it follows that $M_{\mathcal{H}}$ has a supremum given by $M_s=R_s/2$.
	Affirmatively, spacetime regularity is possible even when $\sup M_s= +\infty$.
	A concrete example—replacing $F$ in Eq.\ref{gtf} with $F = (f - h + nk)^2$, setting $m=0,nk>1/(a-L)$, and keeping other relations—yields finite curvature invariants ($R$, $R_{ab}R^{ab}$, $R_{abcd}R^{abcd}$).

	This spacetime is globally hyperbolic despite the absence of conformal null boundaries.
	The horizon behavior as $r \to 0$ is: spacelike for $k > 1/(2a(a-L)^2)$, null for $k = 1/(2a(a-L)^2)$, and timelike for $k < 1/(2a(a-L)^2)$.
	Its Penrose-Carter diagram (Fig.~\ref{f4}) illustrates the process whereby a GSLBH comes to trap the entire future spacetime.
	The horizonless wormhole case is omitted, as it is given by the simple substitution of $kT^2$ in $h$ with $lT^4$.
	
	\section{Results and Discussion}
	In conclusion, this work establishes a novel framework for generating arbitrary spherically symmetric metrics through the coupling of two scalar fields with an electromagnetic field.
	By specializing this framework to a class of analytically tractable, non-stationary spacetimes, we have systematically investigated the dynamical evolution of black holes and wormholes.
	Our central finding is a three-fold classification for the ultimate fate of such spacetimes, with the globally trapped future emerging as a particularly significant new possibility.
	Our findings open new avenues for modeling the evolution of non-singular spacetime objects.
	
	It was first pointed out by Hayward \cite{hayward1999dynamic} that black holes and wormholes can transform into each other.
	In our example (Fig.~\ref{f4}), it is shown that a regular black hole constitutes a collection of throats bounded by spacelike marginally trapped horizons, without involving a topology change (as in the Bardeen black hole \cite{bardeen1968non} and its variants \cite{fan2016construction}, which contain a de Sitter core) or a “black-bounce” core (as in the Simpson-Visser black hole \cite{simpson2019black} and its variants \cite{lobo2021novel}).
	
	Hence, we propose that a regular black hole may contain an unattainable minimal throat---a minimal-radius throat that is causally unreachable---and refer to it henceforth as the Endless Throat.
	
	One might question that while spacetimes like Bardeen's and Simpson-Visser's are constructed from the Schwarzschild black hole and are static in their domains of outer communication, our present examples, constructed from the Ellis-Bronnikov wormhole for greater controllability and simplicity, lacks such a region (and consequently lacks the corresponding Killing field) and cannot reduce to the Schwarzschild black hole in an appropriate limit.
	This could be seen as diminishing the persuasiveness of our model.
	In response, we contend that it is possible to obtain a regular black hole that (i) reduces to the Schwarzschild black hole in an appropriate limit, (ii) contains the requisite Killing field and a bifurcate Killing horizon, and (iii) retains an Endless Throat.
	The pursuit of such a configuration remains a central and open direction for future work.
	This perspective is further supported by the naturalness of the Endless Throat concept: it generalizes the Schwarzschild interior, where all inward causal trajectories terminate at a singular endpoint, by replacing that endpoint with a causally unattainable, minimal throat—thus maintaining the inescapable nature of the black hole while eliminating the singularity.
	
\section*{References}
\bibliography{db-9}

\providecommand{\newblock}{}
\begin{thebibliography}{10}
\expandafter\ifx\csname url\endcsname\relax
  \def\url#1{{\tt #1}}\fi
\expandafter\ifx\csname urlprefix\endcsname\relax\def\urlprefix{URL }\fi
\providecommand{\eprint}[2][]{\url{#2}}

\bibitem{schwarzschild1916gravitationsfeld}
Schwarzschild K 1916 {\em Sitzungsberichte der k{\"o}niglich preu{\ss}ischen
  Akademie der Wissenschaften zu Berlin\/}  424--434

\bibitem{wald2024general}
Wald R~M 2024 {\em General relativity\/} (University of Chicago press)

\bibitem{penrose1965gravitational}
Penrose R 1965 {\em Physical Review Letters\/} {\bf 14} 57

\bibitem{hawking1967occurrence}
Hawking S~W 1967 {\em Proceedings of the Royal Society of London. Series A.
  Mathematical and Physical Sciences\/} {\bf 300} 187--201

\bibitem{hawking1970singularities}
Hawking S~W and Penrose R 1970 {\em Proceedings of the Royal Society of London.
  A. Mathematical and Physical Sciences\/} {\bf 314} 529--548

\bibitem{bardeen1968non}
Bardeen J 1968 Non-singular general relativistic gravitational collapse {\em
  Proceedings of the 5th International Conference on Gravitation and the Theory
  of Relativity\/} p~87

\bibitem{ayon1998regular}
Ayon-Beato E and Garcia A 1998 {\em Physical review letters\/} {\bf 80} 5056

\bibitem{ayon2000bardeen}
Ayon-Beato E and Garc{\i}a A 2000 {\em Physics Letters B\/} {\bf 493} 149--152

\bibitem{bronnikov2000comment}
Bronnikov K 2000 {\em Physical review letters\/} {\bf 85} 4641

\bibitem{bronnikov2001regular}
Bronnikov K~A 2001 {\em Physical Review D\/} {\bf 63} 044005

\bibitem{burinskii2002new}
Burinskii A and Hildebrandt S~R 2002 {\em Physical Review D\/} {\bf 65} 104017

\bibitem{fan2016construction}
Fan Z~Y and Wang X 2016 {\em Physical Review D\/} {\bf 94} 124027

\bibitem{rodrigues2022bardeen}
Rodrigues M~E, de~S~Silva M~V and Vieira H~A 2022 {\em Physical Review D\/}
  {\bf 105} 084043

\bibitem{dariescu2022kiselev}
Dariescu M~A, Dariescu C, Lungu V and Stelea C 2022 {\em Physical Review D\/}
  {\bf 106} 064017

\bibitem{bronnikov2023regular}
Bronnikov K~A 2023 Regular black holes sourced by nonlinear electrodynamics
  {\em Regular Black Holes: Towards a New Paradigm of Gravitational Collapse\/}
  (Springer) pp 37--67

\bibitem{moreno2003stability}
Moreno C and Sarbach O 2003 {\em Physical Review D\/} {\bf 67} 024028

\bibitem{novello2000geometrical}
Novello M, De~Lorenci V, Salim J and Klippert R 2000 {\em Physical Review D\/}
  {\bf 61} 045001

\bibitem{novello2000singularities}
Novello M, Bergliaffa S~P and Salim J 2000 {\em Classical and Quantum
  Gravity\/} {\bf 17} 3821

\bibitem{ma2014corrected}
Ma M~S and Zhao R 2014 {\em Classical and Quantum Gravity\/} {\bf 31} 245014

\bibitem{bokulic2021black}
Bokuli{\'c} A, Smoli{\'c} I and Juri{\'c} T 2021 {\em Physical Review D\/} {\bf
  103} 124059

\bibitem{simpson2019black}
Simpson A and Visser M 2019 {\em Journal of Cosmology and Astroparticle
  Physics\/} {\bf 2019} 042

\bibitem{lobo2021novel}
Lobo F~S, Rodrigues M~E, Silva M~V~d~S, Simpson A and Visser M 2021 {\em
  Physical Review D\/} {\bf 103} 084052

\bibitem{bronnikov2022black}
Bronnikov K 2022 {\em Physical Review D\/} {\bf 106} 064029

\bibitem{bronnikov2022field}
Bronnikov K~A and Walia R~K 2022 {\em Physical Review D\/} {\bf 105} 044039

\bibitem{canate2022black}
Ca{\~n}ate P 2022 {\em Physical Review D\/} {\bf 106} 024031

\bibitem{hayward1999dynamic}
Hayward S~A 1999 {\em International Journal of Modern Physics D\/} {\bf 8}
  373--382

\bibitem{shinkai2002fate}
Shinkai H~a and Hayward S~A 2002 {\em Physical Review D\/} {\bf 66} 044005

\bibitem{hayward2009wormhole}
Hayward S~A 2009 {\em Physical Review D—Particles, Fields, Gravitation, and
  Cosmology\/} {\bf 79} 124001

\bibitem{bronnikov2007regular}
Bronnikov K, Dehnen H and Melnikov V 2007 {\em General Relativity and
  Gravitation\/} {\bf 39} 973--987

\bibitem{Hayward:2005gi}
Hayward S~A 2006 {\em Phys. Rev. Lett.\/} {\bf 96} 031103 (\textit{Preprint}
  \eprint{gr-qc/0506126})

\bibitem{simpson2019vaidya}
Simpson A, Martin-Moruno P and Visser M 2019 {\em Classical and Quantum
  Gravity\/} {\bf 36} 145007

\bibitem{lobo2020dynamic}
Lobo F~S, Simpson A and Visser M 2020 {\em Physical Review D\/} {\bf 101}
  124035

\bibitem{yang2021trapping}
Yang J and Huang H 2021 {\em Physical Review D\/} {\bf 104} 084005

\bibitem{liang2025regular}
Liang Y~b and Li H~R 2025 {\em arXiv preprint arXiv:2506.13676\/}

\bibitem{ashtekar2014geometry}
Ashtekar A 2014 {\em arXiv preprint arXiv:1409.1800\/}

\end{thebibliography}

\end{document}